# A corrected open boundary framework for lattice Boltzmann immiscible pseudopotential models

View Online

Export Citation

CrossMark

Yizhong Chen (陈怡忠),[1,2] Huajie Pan (潘铧杰),[1,2] Zhibin Wang (王智彬),[1,2,a)] Dongliang Li (李栋梁),[3,a)] 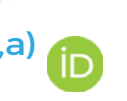 and Ying Chen (陈颖)[1,2]

## AFFILIATIONS

[1]School of Material and Energy, Guangdong University of Technology, Guangzhou 510006, China
[2]Guangdong Provincial Key Laboratory of Functional Soft Matter, Guangzhou 510006, China
[3]BGI Research, Shenzhen 518083, China

[a)]**Authors to whom correspondence should be addressed:** wangzhibin@gdut.edu.cn. **Tel.:** +86 020-39322570 and lidongliang@genomics.cn

## ABSTRACT

The pseudopotential lattice Boltzmann method is a prominent approach for simulating multiphase flows, valued for its physical intuitiveness and computational tractability. However, existing immiscible pseudopotential methods for modeling dynamic multi-component immiscible fluid systems involving open boundaries face persistent challenges, notably the influence of spurious currents on interface formation and breakup, as well as the effects of inlet and outlet boundary configurations on simulation stability. Therefore, this paper proposes a corrected open boundary framework based on multiple-relaxation-time for the immiscible pseudopotential model. Our method includes three key improvements: (1) For the accurate recovery of macroscopic quantities at the inlet boundary, correction coefficients are introduced to reconstruct the distribution function; (2) Based on real-time mass flow rates at the inlet and outlet, the outlet boundary velocity is adjusted to ensure global mass conservation in the computational domain; (3) The relaxation coefficient related to numerical stability is adjusted based on the viscosity of two-phase fluids to reduce spurious currents. To validate the reliability of the proposed corrected method, four benchmark cases were simulated: Laplace tests and Taylor deformation, two-phase Poiseuille flow, migration of droplets in microchannels, as well as droplet generation in T-shaped and co-flow devices. The results demonstrate that the corrected method reduced the average spurious currents at the phase interface by 65.8% and controlled the average mass deviation of the fluid system at around 3.5%. In addition, the morphology of the droplets differs by less than 5% compared to the benchmark examples and experiments.

## NOMENCLATURE

|  |  |
| --- | --- |
| $Ca$ | Capillary number |
| $c_s^2$ | Lattice speed of sound |
| $D$ | Diameter (lu) |
| $\boldsymbol{e}$ | Discrete velocity |
| $\boldsymbol{F}$ | Total force (mu lu/ts$^2$) |
| $\boldsymbol{F}_{ads}$ | Adhesive forces between fluid and solid, lattice units (mu lu/ts$^2$) |
| $\boldsymbol{F}_f$ | Intermolecular forces between two fluids, lattice units (mu lu/ts$^2$) |
| $\bar{\boldsymbol{F}}$ | External forces term |
| $f$ | Density distribution function |
| $f^{eq}$ | Equilibrium density distribution function |
| $G_w$ | Strength coefficient of interaction between solid and fluid |
| $G_{12}$ | Strength coefficient of interaction between two fluids |
| $M$ | Viscosity ratio |
| $p$ | Pressure [mu/(ts$^2$ lu)] |
| $Q$ | Flux (lu$^3$/ts) |
| $R$ | Radius (lu) |
| $\boldsymbol{S}$ | Diagonal relaxation matrix |
| $\boldsymbol{T}$ | Transformation orthogonalization matrix |
| $\boldsymbol{u}$ | Velocity, lattice units (lu/ts) |
| $\boldsymbol{u}^{eq}$ | Common velocity, lattice units (lu/ts) |
| $W$ | Width (lu) |
| $w$ | Weight coefficient |
| $x$ | Horizontal direction of Cartesian coordinate system |
| $y$ | Vertical direction of Cartesian coordinate system |
| $z$ | Height direction of Cartesian coordinate system |

### Greek symbols

|  |  |
| --- | --- |
| $\alpha$ | Distribution functions correction coefficient |
| $\gamma$ | Surface tension coefficient |
| $\delta_t$ | Time step, lattice units (ts) |
| $\varepsilon$ | Relative error |

$\lambda$ Velocity correction coefficient
$\rho$ Density, lattice units (mu/lu$^3$)
$\upsilon$ Kinematic viscosity, lattice units (lu$^2$/ts)
$\psi$ Pseudopotential function

### Superscripts/subscripts

$a$ Analytical solution
$c$ Continuous phase
$d$ Dispersed phase
$i$ Lattice directions
$k$ Index of different fluid components

## I. INTRODUCTION

The complex dynamic multiphase flows have always been an interesting issue in many scientific and industrial applications, especially involving open boundaries,[1–3] such as inlet and outlet. To investigate the mechanisms of these complex flow phenomena, numerous researchers have employed various multiphase flow numerical methods for simulation, such as Volume of Fluid,[4] Level-Set,[5] and Phase-Field[6] methods. Although these numerical methods have achieved good results under macroscopic conditions, they still face challenges under microscale conditions, including numerical dissipation at the phase interface and failure of the continuum assumption, leading to deviations from actual microscopic mechanisms.[7–10] Therefore, a new numerical simulation method, lattice Boltzmann method (LBM) which is based on kinetic theory, has garnered increasing attention in multiphase flow simulations. The kinetic nature of LBM provides distinct advantages in interfacial handing through automatic phase interface maintenance, effectively circumventing the intricate interface tracking procedures required by conventional numerical methods and is inherently free from the continuum assumption.[11,12] Among the diverse LBM multiphase methods developed over the past decades,[13–15] the pseudopotential model distinguishes itself through its elegant physical interpretation and computational simplicity. This innovative approach generates interphase interactions forces through pseudopotential gradient that mimics long-range attractive and short-range repulsive molecular interactions, thereby enabling spontaneous phase separation and capable of automatically modeling dynamic contact angles, and these features have led to its widespread adoption in the phenomenon of various complex dynamic multiphase flows.[16]

For many flows of interest, the ideal domain should be very large or infinitely long in one or more dimensions. In practical computations, however, it is necessary to artificially truncate the domain to a finite field due to computer resources. Thus, a proper open boundary treatment is a critical ingredient of numerical simulations. In pseudopotential LBM implementations, precise open boundaries require carefully designed inlet and outlet schemes. For the inlet boundary, the non-equilibrium (NEQ) bounce-back scheme proposed by Zou and He[17] is widely used in 2D models due to its third-order accuracy and mass conservation. However, in 3D, it suffers from computational complexity and difficulties in handling corners, often reducing accuracy or causing divergence, especially with complex the inlet boundary.[18–21] This limits its cross-dimensional versatility. In contrast, the second-order NEQ extrapolation scheme by Zhao-Li *et al.*[22] maintains consistent ease of implementation across dimensions, requiring only a simple linear extrapolation to obtain boundary distribution functions. This significant advantage holds considerable appeal for us, particularly as 3D simulations are increasingly adopted for their ability to more accurately represent real-world flows. However, the interpolation property of this scheme makes it difficult for the distribution function on the boundary to accurately recover the given macroscopic physical quantities, resulting in a decrease in simulation accuracy.[23,24] To address this issue and inspired by Ju *et al.*,[25] this paper proposes a distribution function correction method: on the basis of the original NEQ scheme, a correction coefficient is further introduced to reconstruct the distribution function on the boundary, ensuring accurate implementation of the given macroscopic quantities at the boundary, eliminating interpolation errors, and improving simulation accuracy. For the outlet boundary, the truncation of the phase interface as it exits the domain can induce spurious numerical artifacts in the adjacent region, thereby compromising the computational stability of the simulation.[26,27] To improve these issues, Lou *et al.*[28] proposed an outflow boundary scheme, which has since been widely adopted in complex multiphase systems such as boiling and fuel cells.[14,29,30] However, this paper observed that under conditions of prolonged and continuous interfacial disturbances, such as droplet generation, the outflow scheme exhibits significant mass non-conservation problems. To address the problem, under the framework of single component vapor–liquid two-phase model, Wang *et al.*[31] developed a technique known as artificial condensation based on the equation of state describing phase change, which promotes the dissipation of the phase interface prior to its arrival at the outlet boundary, thereby effectively circumventing instabilities induced by interfacial perturbations near the outflow region. Additionally, under the frame of color-gradient model, Zong *et al.*[32] had reconstructed the distribution function at the outlet based on the transport equation of phase fraction, to maintain conservation of mass. However, the immiscible pseudopotential model described in this article is difficult to directly apply the methods of the aforementioned scholars to this model due to the absence of the equation of state to describe phase change and a governing equation defined by the phase fraction in its algorithm system. Actually, this paper has found that the non-conservation of mass within the computational domain stemmed from an imbalance between the inlet and outlet mass flow rates. Therefore, inspired by Tong *et al.*,[33] this paper proposes a velocity correction method: on the basis of the original outflow scheme, a velocity correction coefficient is further introduced to correct the velocity on the outlet boundary, ensuring that the mass flow rate at the inlet and outlet is in an equilibrium state to satisfy mass conservation.

In addition, suppressing spurious currents is another challenge to improve numerical accuracy and stability in immiscible pseudopotential method. It is a nonphysical phenomenon at the phase interface of non-zero curvature, and excessive spurious currents usually lead to the formation and breakup of the phase interface deviating from the reality.[34–36] To mitigate spurious currents, researchers have proposed different improved collision operators[37–39] and higher-order interaction force formats.[40] Considering computational cost and the attenuation level of spurious currents, this paper adopts Multi relaxation time (MRT) collision operator and the eighth-order interaction force format. In addition to the methods mentioned above, this paper also found that adjusting the relaxation coefficient $s_\varepsilon$ that affects numerical stability[41] based on the viscosity of two-phase fluids can also suppress spurious currents. Compared to the spurious currents before adjustment, this method can reduce it by two orders of magnitude, which highlights the method's exceptional capability in addressing immiscible flows. Moreover, this adjustment method is based on the physical characteristics of fluid systems beyond traditional empirical adjustment methods, making the parameter optimization of the model more instructive and achieving maximum simulation fidelity.

Critical analysis of existing methodologies reveals persistent challenges in pseudopotential-based immiscible multiphase simulations: (1) Original non-equilibrium extrapolation scheme cannot accurately recover macroscopic quantities at the boundary, resulting in a decrease in computational accuracy; (2) the global mass in the computational domain is not conserved when using the original outflow scheme; and (3) larger spurious currents usually cause the formation and breakup of the phase interface to deviate from reality. The advancement we proposed improves these limitations by correcting existing schemes, specifically by introducing correction coefficients to reconstruct the distribution function, in order to accurately recover the boundary's macroscopic quantities and adjusting the boundary's velocity based on real-time mass flow rates to ensure global mass conservation. For the problem of spurious currents, the relaxation coefficient related to numerical stability is adjusted based on the viscosity of two-phase fluids to suppress it. Implementation validation employs a hierarchical verification protocol: (1) Classic Laplace and Taylor deformation verification, (2) Quantitative benchmarking against analytical solutions and prior numerical studies, and (3) Dynamic multiphase scenarios. Theoretical significance emerges from its synthetic resolution to the inherent conflicts among boundary accuracy, mass conservation, and spurious currents in multiphase flow simulation. Practical implications extend to high-stability simulations in microscale droplet dynamics where conventional methods exhibit prohibitive numerical diffusion.

## II. PSEUDOPOTENTIAL MULTIPHASE MODEL OF LBM

### A. Theoretical framework

LBM operates at the mesoscopic scale, resolving fluid dynamics through discrete distribution functions $\boldsymbol{f}_i^k(x,\,t)$ that evolve via collision and streaming processes on a structured lattice. For immiscible two-phase systems, the evolution equations for each component $k\in\{1,\ 2\}$. This kinetic-theory-based approach inherently captures interfacial dynamics in multiphase systems without explicit interface tracking, making it particularly suited for immiscible fluid simulations.

For immiscible two-phase flows, the pseudopotential model employs dual sets of distribution functions $(\boldsymbol{f}_i^1, \boldsymbol{f}_i^2)$ to represent distinct fluid components. Each component adheres to the discrete Boltzmann equation as follows:

$$\boldsymbol{f}_i^k(\boldsymbol{x}, t+\Delta t) - \boldsymbol{f}_i^k(\boldsymbol{x}, t) = -(\boldsymbol{T}^{-1}\boldsymbol{S}\,\boldsymbol{T})\left(\boldsymbol{f}_i^k(\boldsymbol{x}, t) - \boldsymbol{f}_i^{k(eq)}(\boldsymbol{x}, t)\right) + \left(\boldsymbol{T}^{-1}\left(\boldsymbol{I} - \frac{\boldsymbol{S}}{2}\right)\boldsymbol{T}\right)\bar{\boldsymbol{F}}_i^k(\boldsymbol{x}, t), \tag{1}$$

$$\boldsymbol{f}_i^k(\boldsymbol{x}+\boldsymbol{e}_i\Delta t, t+\Delta t) = \boldsymbol{f}_i^k(\boldsymbol{x}, t+\Delta t). \tag{2}$$

where $\boldsymbol{f}_i^k$ and $\boldsymbol{f}_i^{k(eq)}$ are the distribution function and the equilibrium distribution function of component $k \in (1, 2)$ at position $\boldsymbol{x}$ and time $t$, respectively. $\boldsymbol{e}_i$ denotes discrete lattice velocities. $\boldsymbol{T}$ is a transformation orthogonalization matrix.

$$\boldsymbol{T} = \begin{bmatrix} 1 & 1 & 1 & 1 & 1 & 1 & 1 & 1 & 1 \\ -4 & -1 & -1 & -1 & -1 & 2 & 2 & 2 & 2 \\ 4 & -2 & -2 & -2 & -2 & 1 & 1 & 1 & 1 \\ 0 & 1 & 0 & -1 & 0 & 1 & -1 & 1 & 1 \\ 0 & -1 & 0 & 2 & 0 & 1 & -1 & 1 & -1 \\ 0 & 0 & 1 & 0 & -1 & 1 & -1 & -1 & -1 \\ 0 & 0 & -2 & 0 & 2 & 1 & -1 & -1 & 0 \\ 0 & 1 & -1 & 1 & -1 & 0 & 0 & 0 & 1 \\ 0 & 0 & 0 & 0 & 0 & 1 & 1 & 1 & 1 \end{bmatrix}. \tag{3}$$

$\boldsymbol{S}$ is a diagonal relaxation matrix which can be expressed as

$$\boldsymbol{S} = \mathrm{diag}[s_\rho, s_e, s_\varepsilon, s_j, s_q, s_j, s_q, s_\vartheta, s_\vartheta]. \tag{4}$$

$s_\rho$ and $s_j$ are conserved quantities with a value of 1; $s_e$, $s_\varepsilon$, and $s_q$ are adjustable relaxation parameters that do not affect macroscopic fluid dynamics but are related to the stability of the model. Therefore, the selection of appropriate relaxation parameters can effectively improve the stability of the model. Based on this principle, this paper adjusts the parameters $s_\varepsilon$ according to the kinematic viscosity of the two-phase fluid as a benchmark to suppress spurious currents at the phase interface, thereby enhancing the stability of the model and for specific details, please refer to Sec. III A; $s_\upsilon$ corresponds to momentum diffusion and its value is calculated by fluid's kinematic viscosity ($\upsilon_k$).

$$\upsilon_k = \frac{1}{3}\left(\frac{1}{s_\upsilon} - \frac{1}{2}\right)\delta_t. \tag{5}$$

The equilibrium distribution function $\boldsymbol{f}_i^{k(eq)}$ is written as

$$\boldsymbol{f}_i^{k(eq)} = \rho_k\omega_i\left[1 + \frac{(\boldsymbol{e}_i\cdot\boldsymbol{u^{eq}})}{{c_s}^2} + \frac{(\boldsymbol{e}_i\cdot\boldsymbol{u^{eq}})^2}{2{c_s}^2} - \frac{|\boldsymbol{u^{eq}}|^2}{2{c_s}^2}\right]. \tag{6}$$

$\omega_i$ is the weight factor with $\omega_0 = 4/9$, $\omega_{1-4} = 1/9$, $\omega_{5-8} = 1/36$, and $c_s = 1/\sqrt{3}$. The combined velocity $\boldsymbol{u^{eq}}$ is calculated by

$$\boldsymbol{u^{eq}} = \sum\nolimits_{k=1}^{2}\rho_k\boldsymbol{u}_k \Big/ \sum\nolimits_{k=1}^{2}\rho_k. \tag{7}$$

where $\rho_k$ and $\boldsymbol{u}_k$ are density and velocity of component $k$ and are calculated as

$$\rho_k = \sum\nolimits_{i=0}^{8}\boldsymbol{f}_i^k, \qquad \rho_k\boldsymbol{u}_k = \sum\nolimits_{i=0}^{8}\boldsymbol{e}_i\boldsymbol{f}_i^k + \frac{\Delta t}{2}\boldsymbol{F}^k. \tag{8}$$

The force term is defined as

$$\bar{\boldsymbol{F}}_i^k = \omega_i\left[\frac{\boldsymbol{e}_i - \boldsymbol{u^{eq}}}{{c_s}^2} + \frac{(\boldsymbol{e}_i\cdot\boldsymbol{u^{eq}})\boldsymbol{e}_i}{{c_s}^4}\right]\cdot\boldsymbol{F}^k, \tag{9}$$

where $\boldsymbol{F}^k$ is the total force acting on component $k$ including the fluid–fluid interaction and solid–fluid interaction force

$$\boldsymbol{F}_f^k(x) = -G_{12}\psi_k(x)\sum\nolimits_{i=0}^{25} w\left(|\boldsymbol{e}_i|^2\right)\psi_k(x+\boldsymbol{e}_i)\boldsymbol{e}_i, \tag{10}$$

$$\boldsymbol{F}_{ads}^k(x) = -G_{w,k}\psi_k(x)\sum\nolimits_{i=0}^{25} w\left(|\boldsymbol{e}_i|^2\right)s(x+\boldsymbol{e}_i\Delta t)\boldsymbol{e}_i, \tag{11}$$

$$G_{w,k}=\begin{cases}G_{w,1}, & \rho^{e}>\delta,\\ g_{1}\left(\rho^{e}\right), & \delta\geq\rho^{e}>0,\\ g_{2}\left(\rho^{e}\right), & 0\geq\rho^{e}>-\delta,\\ G_{w,2}, & \rho^{e}<-\delta.\end{cases} \tag{12}$$

$G_{12}$ is interaction strength coefficient between different components and its value can be adjusted to change surface tension and then achieve phase separation. Different contact angles can be achieved by adjusting $G_{w,k}$, and its specific value is calculated using Eq. (12). The detailed information for this interaction scheme can be found in our previous work.[42] $\psi_k(x)$ is pseudopotential function taken as $\psi_k(x) = \rho_k(x)$ for immiscible pseudopotential and $w\left(|e_i|^2\right)$ represents the weight coefficients for different density gradient direction. Considering the particularity of the inlet and outlet boundary and reducing the spurious currents as much as possible, the interaction between different components is calculated in the 8th-order format, and the corresponding weight coefficients are $w(1)=4/21$, $w(2)=4/45$, $w(4)=1/60$, $w(5)=2/315$, and $w(8)=1/5040$.

## B. Correction for inlet and outlet boundary

Considering the simplicity and universality of handling inlet velocity boundaries, this paper proposes a corrected NEQ extrapolation scheme. This method is based on the original scheme and then introduces correction coefficients to correct the distribution function. For the original scheme, the basic idea is to divide the unknown distribution function on the inlet boundary nodes into two parts: equilibrium and NEQ. The equilibrium state is obtained based on the definition of boundary conditions, while the NEQ state is determined by extrapolation from neighbor fluid node. Taking a 2D model as an example, as shown in Fig. 1, if the fluid flows into the fluid domain in a horizontal direction, after the process of migration, the unknown distribution function at the inlet boundary is $f_1^k, f_5^k, f_8^k$; conversely, if the fluid flows into the fluid domain in a vertical direction, the unknown distribution function at the inlet boundary is $f_4^k, f_7^k, f_8^k$. Subsequently, the distribution function to be solved is calculated through the original NEQ extrapolation scheme, as shown in Eqs. (13) and (14), and the correction coefficient for the distribution function is precisely used to process the distribution function obtained from the aforementioned solution. To avoid excessive redundancy, the subsequent discussion about corrected coefficient in this paper will be analyzed based on the scenario of horizontal fluid inflow.

$$\boldsymbol{f}_i^k(x_1)=\boldsymbol{f}_i^{k(eq)}\left(\rho_k(x_1),\boldsymbol{u}^{\boldsymbol{eq}}(x_1)\right)+\left[\boldsymbol{f}_i^k(x_2)-\boldsymbol{f}_i^{k(eq)}\left(\rho_k(x_2),\boldsymbol{u}^{\boldsymbol{eq}}(x_2)\right)\right]\quad i=1,\,5,\,8, \tag{13}$$

$$\boldsymbol{f}_i^k(y_{\mathrm{N}})=\boldsymbol{f}_i^{k(eq)}\left(\rho_k(y_{\mathrm{N}}),\boldsymbol{u}^{\boldsymbol{eq}}(y_{\mathrm{N}})\right)+\left[\boldsymbol{f}_i^k(y_{\mathrm{N}-1})-\boldsymbol{f}_i^{k(eq)}\times\left(\rho_k(y_{\mathrm{N}-1}),\boldsymbol{u}^{\boldsymbol{eq}}(y_{\mathrm{N}-1})\right)\right]i=4,\,7,\,8. \tag{14}$$

Although the original scheme is easy to implement and theoretically has a second-order convergence accuracy, this paper found that when the original scheme is directly applied to the immiscible pseudopotential model, its stability is not as good as that of the single-phase flow model. We believe that this is because two-phase fluid systems are more sensitive to errors compared to single-phase systems, especially during long-term iterative processes. This results in the real flow rate of the inlet not being equal to the given flow rate, causing deviations in the simulation results. Considering this problem and inspired by Ju *et al.*[25] this paper reconstructs the distribution function pointing to the fluid domain ($f_1^k, f_5^k, f_8^k$) on the inlet boundary by introducing correction coefficients, thereby accurately recovering the macroscopic quantities defined on the inlet boundary. The details of implementation are shown below

$$\boldsymbol{f}_i^{*k}=\boldsymbol{f}_i^{k}+\alpha_i\; i=1,5,8, \tag{15}$$

$$\boldsymbol{u}_k^{*}=\left(\sum\nolimits_{i=0}^{8}\boldsymbol{e}_i\boldsymbol{f}_i^k+\frac{\Delta t}{2}\boldsymbol{F}^k\right)\Big/\rho_k+e_1\alpha_1+e_5\alpha_5+e_8\alpha_8, \tag{16}$$

$$e_1\alpha_1+e_5\alpha_5+e_8\alpha_8=\boldsymbol{u}_k^{*}-\boldsymbol{u}_k, \tag{17}$$

$$\left(\sum\nolimits_{i}e_{ix}\alpha_i,\ e_{iy}\alpha_i\right)=\boldsymbol{u}_k^{*}-\boldsymbol{u}_k, \tag{18}$$

$$\alpha_i=\left(\frac{e_{ix}\omega_i}{\sum\nolimits_{i=1,5,8}e_{ix}^2\omega_k},\frac{e_{iy}\omega_i}{\sum\nolimits_{i=1,5,8}e_i^2\omega_k}\right)\cdot\left(\rho_k\boldsymbol{u}_k^{*}-\boldsymbol{u}_k\right). \tag{19}$$

$\boldsymbol{f}_i^{*k}$ is reconstructed distribution function and $\alpha_i$ is introduced to modify $\boldsymbol{f}_i^{k}$ which calculated by Eq. (13). $\boldsymbol{u}_k^{*}$ is the inlet velocity after correcting, $\rho_k$ and $\boldsymbol{u}_k$ are physical quantities given on the inlet boundary. $e_{ix}$ and $e_{iy}$ are the components of $\boldsymbol{e}_i$.

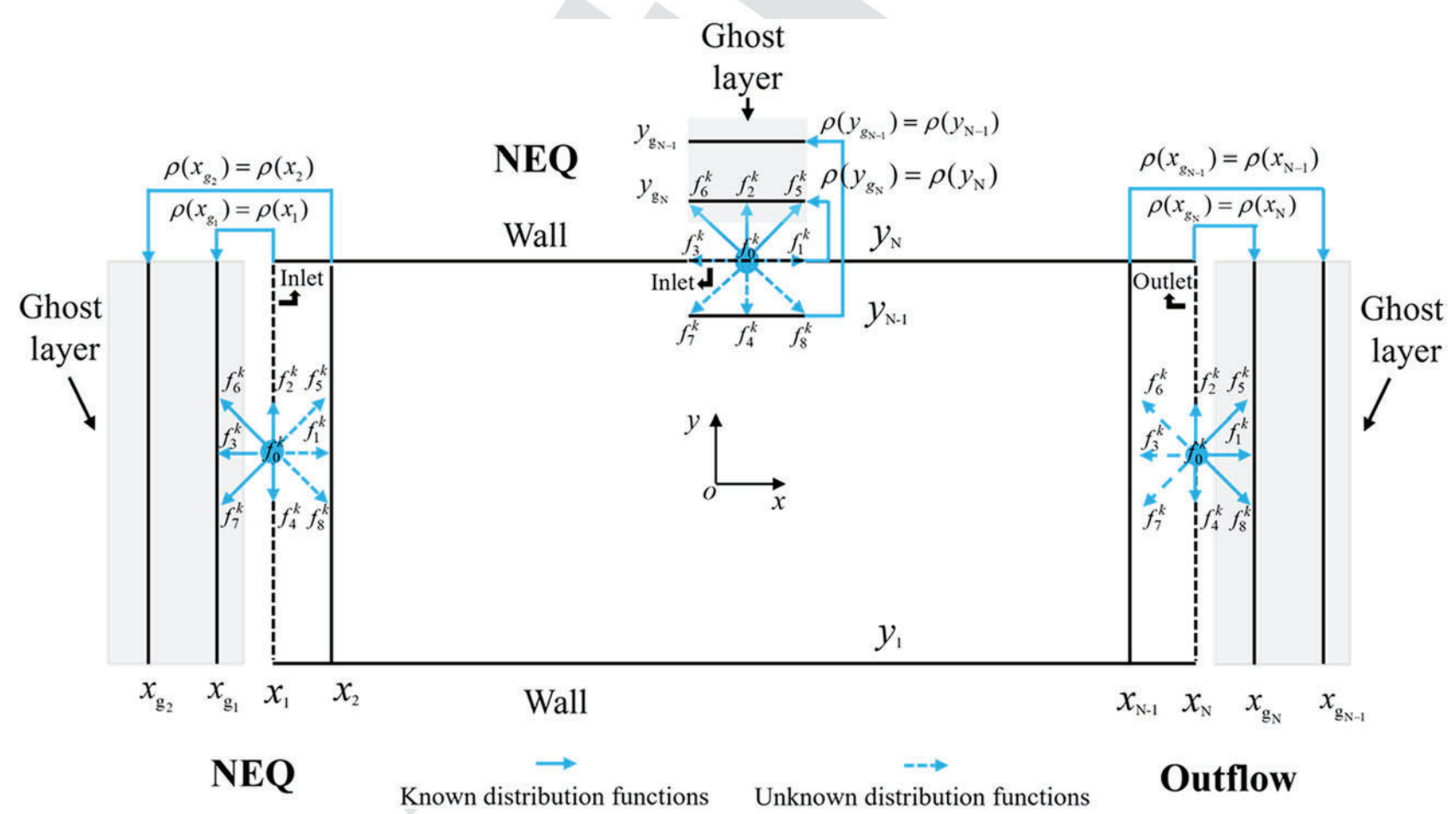

**FIG. 1.** Schematic diagram of the inlet/outlet boundary and ghost layer.

For the outlet boundary, this paper proposes a corrected outflow boundary scheme by introducing a velocity correction coefficient. First, the calculation expression of its distribution function and schematic diagram are shown as Eqs. (20) ad (21) and Fig. 1

$$f_i^k(x_{\mathrm{N}}, j, t+\Delta t) = \frac{f_i^k(x_{\mathrm{N}}, j, t) + \lambda \cdot f_i^k(x_{\mathrm{N}-1}, j, t+\Delta t)}{1+\lambda} \quad i = 3,\ 6,\ 7, \tag{20}$$

$$\lambda = \frac{1}{M}\sum_j u_{k,x}(x_{\mathrm{N}-1}, j, t+\Delta t). \tag{21}$$

$\lambda$ and $M$ are the average combined velocity and the number of nodes at the $x_{\mathrm{N}-1}$, respectively. To overcome the problem of mass non-conservation, this paper is based on real-time flow of inlet and outlet, adjusting velocity at the outlet to ensure mass conservation within the calculation domain and the specific expression as follows:

$$u_{k,x}(x_{\mathrm{N}}, j, t+\Delta t) = \chi \cdot u_{k,x}(x_{\mathrm{N}-1}, j, t+\Delta t), \tag{22}$$

$$\chi = \frac{\sum_{j=1}^{M}\left(\rho_1(x_1,j,t+\Delta t)+\rho_2(x_1,j,t+\Delta t)\right)\cdot u_{k,x}(x_1,j,t+\Delta t)}{\sum_{j=1}^{M}\left(\rho_1(x_{\mathrm{N}-1},j,t+\Delta t)+\rho_2(x_{\mathrm{N}-1},j,t+\Delta t)\right)\cdot u_{k,x}(x_{\mathrm{N}-1},j,t+\Delta t)}, \tag{23}$$

where $\chi$ is the velocity correction coefficient at the outlet boundary. In theory, the global conservation of mass is to ensure that the difference in mass flow between inlet and outlet is zero. Therefore, by calculating the mass flow rate at the inlet and outlet, the outlet velocity is corrected using a correction coefficient $\chi$ to ensure the net mass flow rate in the simulation system is zero.

Finally, for calculating the fluid–fluid interaction force on the inlet/outlet boundary and enhancing the mechanical equilibrium at that location, this paper suggests setting up two additional layers of ghost nodes to calculate the force, as shown in Fig. 1. The density setting at the ghost layer is shown in Eqs. (24) and (25). Implementation of this computational strategy can weaken negative impact caused by anisotropy at the boundary. To illustrate the above modeling process clearly, Fig. 2 shows the calculation flow chart in one computational cycle. For code implementation details, please refer to Appendix A

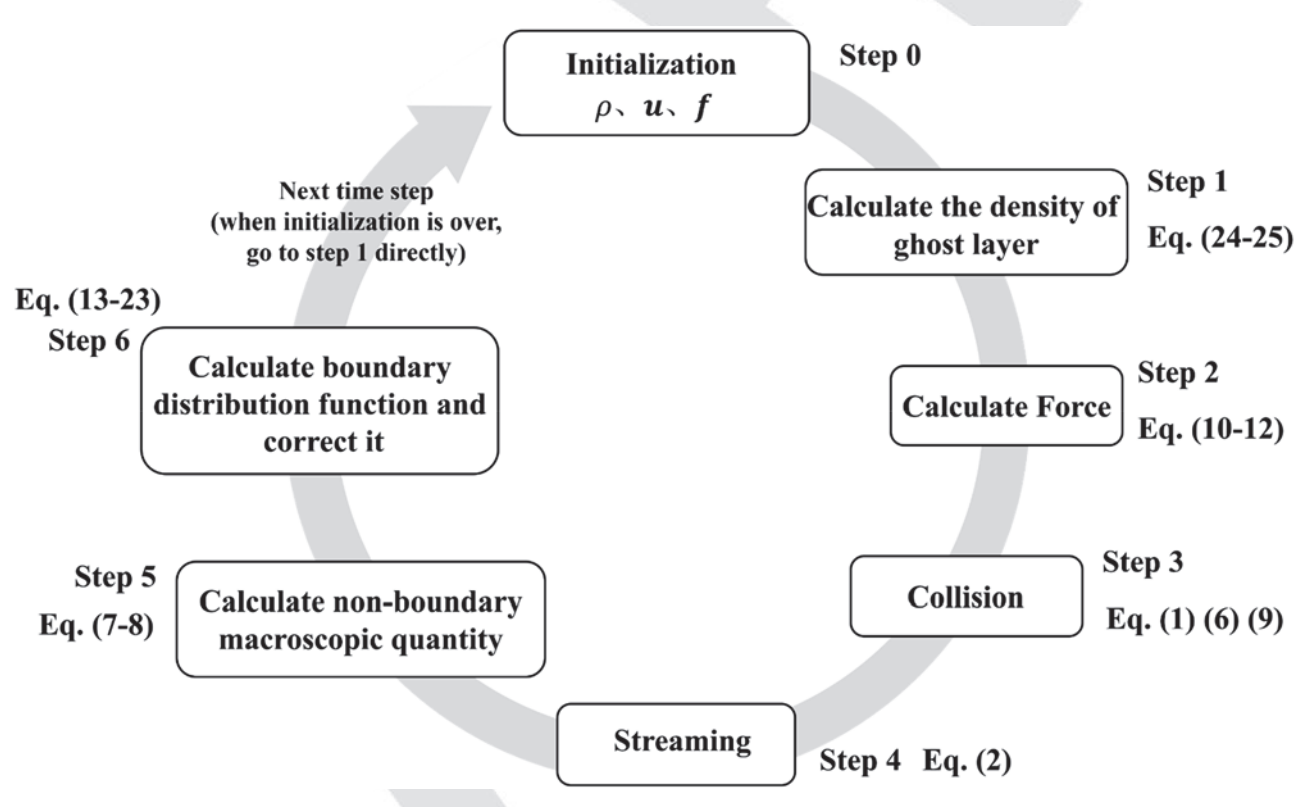

FIG. 2. Flowchart of one computational cycle of the calculation process.

$$\rho_{\mathrm{k}}\left(\mathrm{x}_{\mathrm{g}_1}, \mathrm{j}, \mathrm{t}+\Delta \mathrm{t}\right) = \rho_{\mathrm{k}}(\mathrm{x}_1, \mathrm{j}, \mathrm{t}+\Delta \mathrm{t}), \tag{24a}$$

$$\rho_{\mathrm{k}}\left(\mathrm{x}_{\mathrm{g}_2}, \mathrm{j}, \mathrm{t}+\Delta \mathrm{t}\right) = \rho_{\mathrm{k}}(\mathrm{x}_2, \mathrm{j}, \mathrm{t}+\Delta \mathrm{t}), \tag{24b}$$

$$\rho_k\left(x_{g_{\mathrm{N}}}, j, t+\Delta t\right) = \rho_k(x_{\mathrm{N}}, j, t+\Delta t), \tag{25a}$$

$$\rho_k\left(x_{g_{\mathrm{N}-1}}, , t+\Delta t\right) = \rho_k(x_{\mathrm{N}-1}, j, t+\Delta t). \tag{25b}$$

## III. RESULTS AND DISCUSSION

### A. Laplace tests and Taylor deformation

First, to verify the accuracy of the model in describing the surface tension, we conducted a Laplace verification ($\Delta P = \frac{\gamma}{R}$), where $\Delta P$ is the difference between the pressure inside the droplet and the pressure outside the droplet and $\gamma$ is the surface tension coefficient. Based on this benchmark test, the effects of spurious currents at the phase interface with different surface tension coefficient and viscosity ratios were also further observed. As shown in Fig. 3(a), a static droplet within computational domain of $100 \times 100$, and the droplet radius is $R = 38$. All boundaries are set to be periodic (details of the boundary schemes are provided in Appendix B). The relaxation parameters are chosen as follows:

$$S = [1.0, 1.43, 1.43, 1.0, 1.0, 1.0, 1.2, 1.43, 1.43]. \tag{26}$$

Corresponding to the viscosities of two components $\vartheta_1 = \vartheta_2 = 0.067$ (viscosity ratio: $M = \upsilon_1/\upsilon_2 = 1.0$), the initial densities are set as $\rho_1 = \rho_2 = 1.0$.

According to Fig. 3(b), when the simulation is in a steady state, the pressure difference between the inside and outside of the droplet and the reciprocal of the droplet radius show a good linear relationship with at different $G_{12}$, and the slope of line is the surface tension coefficient, satisfying Laplace's law. This proves that our model can accurately describe the interfacial tension of the two immiscible phases. Since $G_{12}$ significantly affects the spurious currents at the interface, it is necessary to observe the change of the spurious currents in the case of different $G_{12}$ and select an appropriate value to ensure that both sufficient phase separation state and smaller spurious currents can be achieved to make the calculation process more stable. Figure 3(c) shows the maximum spurious currents change at the interface and the corresponding interface thickness ($W_{\mathrm{in}}$) at different $G_{12}$ when $M = 1$, where $u_s = \sqrt{(u_x^{eq})^2 + (u_y^{eq})^2}$, $u_x^{eq}$ and $u_y^{eq}$ represent the combined velocity in the $x$ and $y$ directions, respectively. It is not difficult to find from Fig. 3(c) that with the increase in $G_{12}$, the spurious currents and its rate of change at the interface gradually increase, while the interface width gradually decreases. This means that when $G_{12} > 0.22$, the spurious currents at the interface will exceed $O\ (10^{-2})$ and the interface width will approach 1, it will result in extremely unstable simulations and inaccurate result. On the contrary, as $G_{12}$ decreases, although the magnitude of the spurious currents at the interface decreases to $O\ (10^{-4})$, the width of the interface increases simultaneously, which means that the interface between the two phases of the fluid is gradually blurred and cannot maintain the state of phase separation well. Therefore, considering the above constraints, $G_{12} = 0.18$ is selected in this paper, and the spurious currents and interface width currently are $7.46 \times 10^{-4}$ and four lattices, respectively, ensuring a clear interface between the two-phase fluids and smaller spurious currents. It should be noted that since the evolution of LBM is based on lattice units, according to the dimensionless conversion principle between lattice units and physical units, while ensuring the best accuracy of the phase interface, although the value of $G_{12}$ remains unchanged, different surface tensions can still be simulated by changing the characteristic length or fluid property parameters based on lattice units, as shown in Table I in Appendix C. Furthermore, this paper further observed the influence of different viscosity ratios on the spurious currents at the interface, as shown in Fig. 3(d). With the increase in viscosity ratio, the spurious current at the interface increases gradually. When the viscosity ratio is in the range of 20–150, the average order of spurious currents is $O$ ($10^{-3}$), which means that there are more limitations for the simulation of immiscible fluid flow. To improve the defect, we adjusted a relaxation coefficient ($s_\varepsilon$) that affects numerical stability. As a free parameter, the adjustment of this parameter basically does not affect the evolution of multiphase flow but it will affect the stability of model. For the immiscible pseudopotential model, the density at the interface changes continuously due to the diffusion characteristics of the interface, which means that there is a compressibility effect at the phase interface, so the relative volume of the fluid at the phase interface will change and at this time, the isotropy at the interface will deteriorate and then further amplify the spurious currents. Thus, the adjustment of $s_\varepsilon$ is reflected in the improvement of isotropy at the phase interface, thereby attenuating spurious currents. However, it should be noted that due to the weak compressibility effect of LBM, the relative volume change at the phase interface cannot be completely avoided. When the value of $s_\varepsilon$ is too small, it will also significantly reduce the stability of the model. To balance the stability of the model and weaken the spurious currents at the phase interface, this paper found that satisfactory results can be achieved when $s_\varepsilon$ is adjusted based on the phase with the highest viscosity in the two-phase fluids. Figure 3(d) shows the change of spurious currents after changing $s_\varepsilon$. When the viscosity ratio is in the range of 1–150, the order of spurious velocity always remains $O$ ($10^{-4}$), which further improves the research scope in the field of multiphase flow.

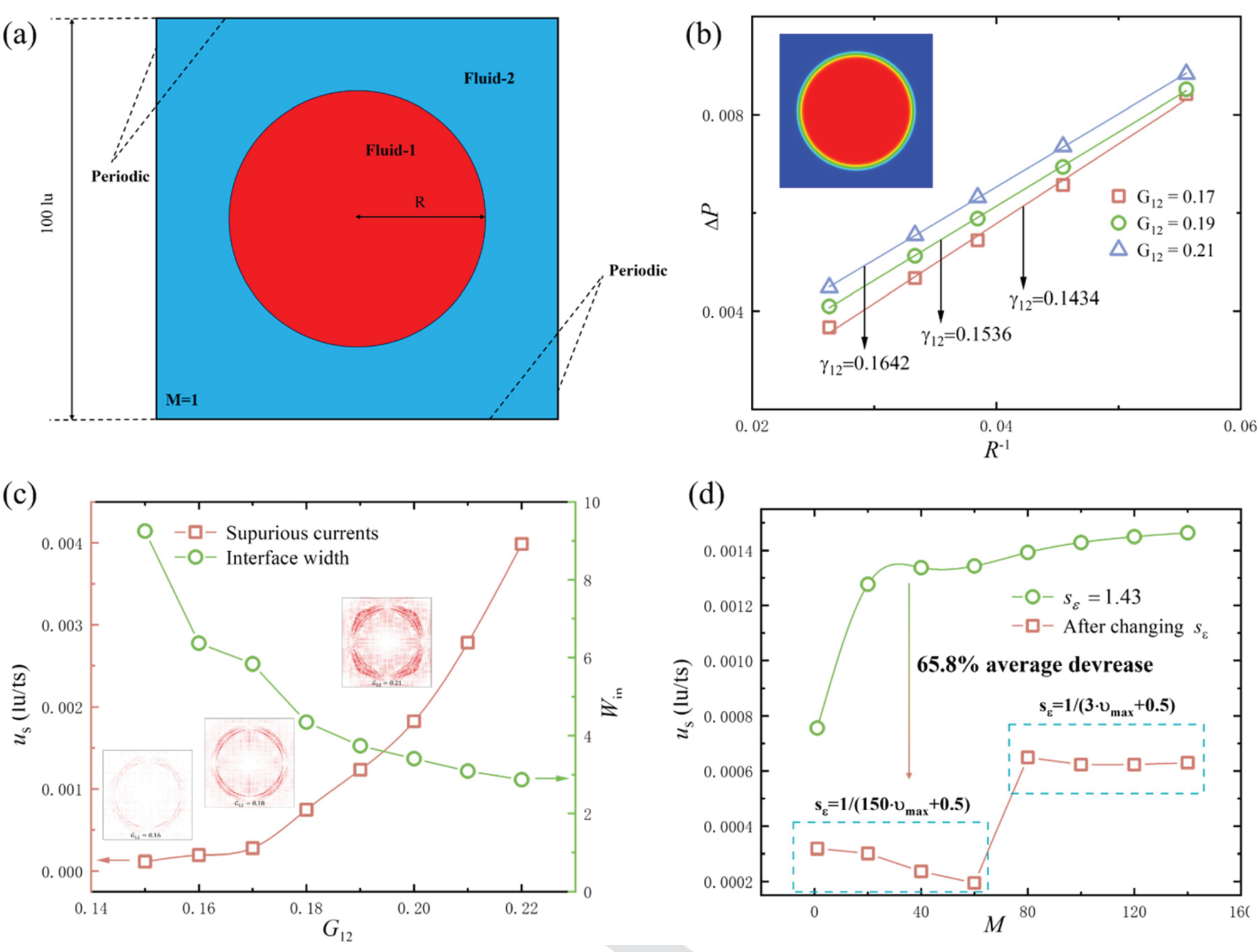

**FIG. 3.** Model and Laplace's test, as well as the change of spurious currents. (a) Schematic diagrams. (b) The change of $\Delta P$ under different $R^{-1}$ and $G_{12}$. (c) Different $u_s$ and $W_{in}$ at various $G_{12}$. (d) The change of $u_s$ at different $M$.

The above case successfully verified the accuracy of adjusting $s_\varepsilon$ on the formation of static phase interfaces. Based on this, Taylor deformation[43] was used in this paper to further test whether the deformation behavior of phase interfaces in dynamic processes is affected by this parameter. A droplet is placed between two shearing plates, which are moving in the opposite directions to obtain a linear shear, as shown in Fig. 4(a). This paper conducted tests based on two viscosity ratios ($M = 30$ and 110), and recorded the degree of droplet deformation (Deformation $= \frac{L-B}{L+B}$, $L$ and $B$ are the half-length and half-width of the droplet, respectively) in a stable state when the error was less than $10^{-6}$. The results are shown in Figs. 4(b) and 4(c). At different capillary numbers ($Ca = \frac{U\upsilon}{\gamma_{12}}$, where $U$ is the velocity of the moving plates), the phase interface of droplets undergoes varying degrees of change, and the degree of deformation increases with the increase in capillary number. In addition, regardless of the viscosity ratio, the degree of deformation at the interface is still linearly related to the capillary number. This indicates that the change in $s_\varepsilon$ does not affect the formation of phase interfaces in the dynamic process.

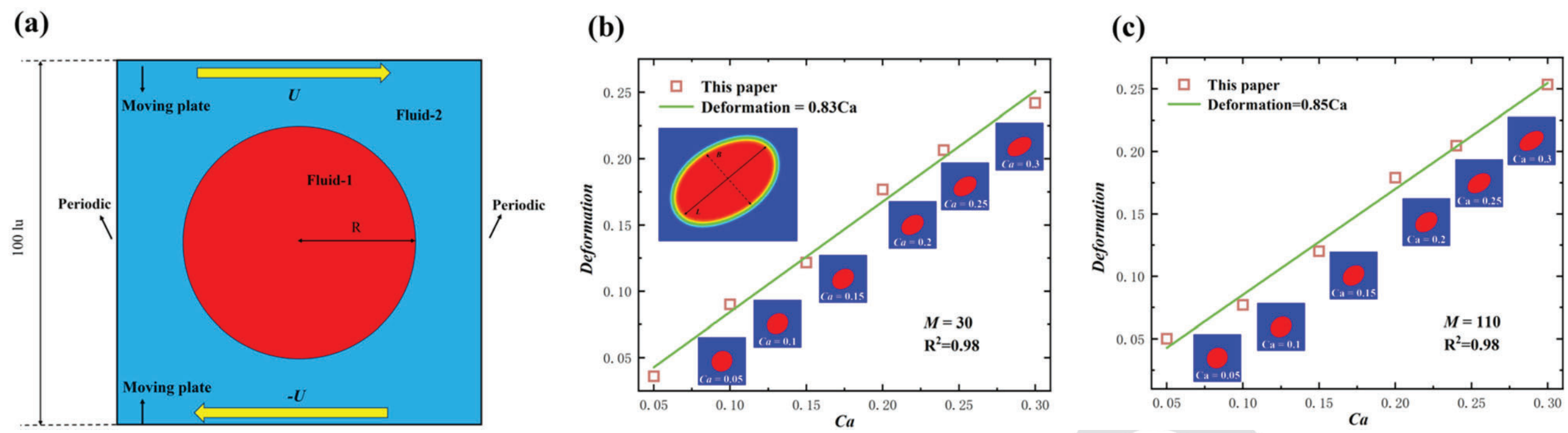

FIG. 4. Model and Taylor deformation under different viscosity ratios. (a) Schematic diagrams. (b) $M = 30$. (c) $M = 110$.

## B. Two-phase Poiseuille flow

Two-phase Poiseuille flow of immiscible fluids is a benchmark for validating numerical algorithms in multiphase systems, as interfacial errors can propagate and distort macroscopic flow behavior, especially when the boundary scheme is not implemented properly or there are differences in fluid properties. To rigorously evaluate the accuracy of the correction methods, we simulate two-phase Poiseuille flow scenarios and compare analytical solutions with numerical solutions. Schematic diagram and analytical solution for validation model are shown in Fig. 5(a) and Eq. (27). The computational domain is a $1000 \times 100$ grid. The upper and lower boundaries are non-slip conditions, and the boundaries on the left and right sides are the inlet and outlet, respectively. Fluid 1 ($\rho_1 = 1.02$, $\upsilon_1 = M * \upsilon_2$, $M = 1/20$, 1, and 20, respectively) is located at $0 \le |y| \le a$ ($a = 25$), and fluid 2 ($\rho_2 = 1.02$, $\upsilon_2 = 0.067$) is located at $a \le |y| \le L$ ($L = 50$). When the absolute difference in velocity between 10 000 time-steps is less than $10^{-6}$, we consider the simulation results to converge.

$$u_x^a(y)_{\text{poiseuille}} = \begin{cases} \dfrac{\boldsymbol{F}_b}{2\upsilon_2\rho_2}\left(L^2 - y^2\right), & a \le |y| \le L \\ \dfrac{\boldsymbol{F}_b}{2\upsilon_1\rho_1}\left(a^2 - y^2\right) + \dfrac{\boldsymbol{F}_b}{2\upsilon_2\rho_2}\left(L^2 - a^2\right), & 0 \le |y| \le a \\ M = 1, \quad \boldsymbol{F}_b = 10^{-7}; \quad M = \dfrac{1}{20} \text{ or } 20, \quad \boldsymbol{F}_b = 5 \times 10^{-6}. & \end{cases} \tag{27}$$

Figure 5(b) compares the normalized velocity profiles from simulations using the correction method with analytical solutions across three viscosity ratios: $M = 1/20$, 1, and 20. The simulated profiles align closely with analytical solutions for the above different flow scenes with relative errors ($\varepsilon$) consistently below 1.52%. This indicates that the correction method can accurately capture flow dynamics under different viscosity ratios and its robustness in diverse interfacial configurations. Additionally, as LBM is a transient numerical method, it often requires a long running time when the system reaches a periodic steady state. If the conservation of system mass is disrupted during this period, it will significantly affect the stability and accuracy of the simulation. Therefore, this paper analyzes the conservation of system mass after implementing the correction method through this case. As shown in Fig. 5(c), before adopting corrected methods, the total mass, which is normalized by the total mass at the initial moment of the system at the current moment, cannot remain stable, and there is a noticeable fluctuation in mass. In addition, the offset of the overall mass of the system increases with the increase in computation time. Therefore, as the computation time further increases, the system will crash. However, after adopting the correction method mentioned in this paper, the total mass of the system remained in a stable state. Although the total mass still exhibits some deviation from the ideal state, the average mass deviation rate $\left(\frac{m_{current} - m_{initial}}{m_{initial}}\right)$ remains below 5%. This indicates that the corrected methods can effectively overcome the issue of mass non-conservation.

## C. Migration of droplets in microchannels

To examine whether the proposed outlet boundary correction method affects droplet behavior compared to the original scheme, a channel flow case with a moving droplet was test under two conditions: unobstructed flow and flow past a rectangular obstacle, as shown in Figs. 6 and 7. First, for the scene without obstacle, the computational domain is a $500 \times 100$ grid. The upper and lower boundaries are non-slip conditions, and the boundaries on the left and right sides are the inlet and outlet boundary, respectively. Droplet ($\rho_1 = 1.02$, $\upsilon_1 = 0.17$) is located at (100, 50), and its radius is 30. The remaining area was filled with fluid 2 ($\rho_2 = 1.02$, $\upsilon_2 = 0.17$). As illustrated in Fig. 6(b), the droplet motion posture before and after correction is basically the same. However, when the droplet flows through the outlet boundary, the deformation morphology of the droplet phase interface predicted by the corrected methods undergoes slight changes compared to the original scheme. We believe this is due to the corrected methods will recalculate the velocity on the outlet boundary, thereby affecting the flow field near the outlet boundary. In fact, this influence is mild, as shown in Figs. 6(c) and 6(d). This paper further compares the instantaneous position ($x_d$) of droplets before the outlet and the degree of deformation of droplets at the outlet quantitatively under two methods. According to Fig. 6(c), it can be observed that the position of the droplet remains consistent at different times when it has not passed through the outlet. When the droplet flows through the outlet, there is a deviation in the position of the droplet using the corrected scheme compared to the original scheme, but the deviation is below 5%. In addition, as shown in Fig. 6(d), there is no significant difference in the deformation of interfaces at different times when the droplet flows through the outlet.

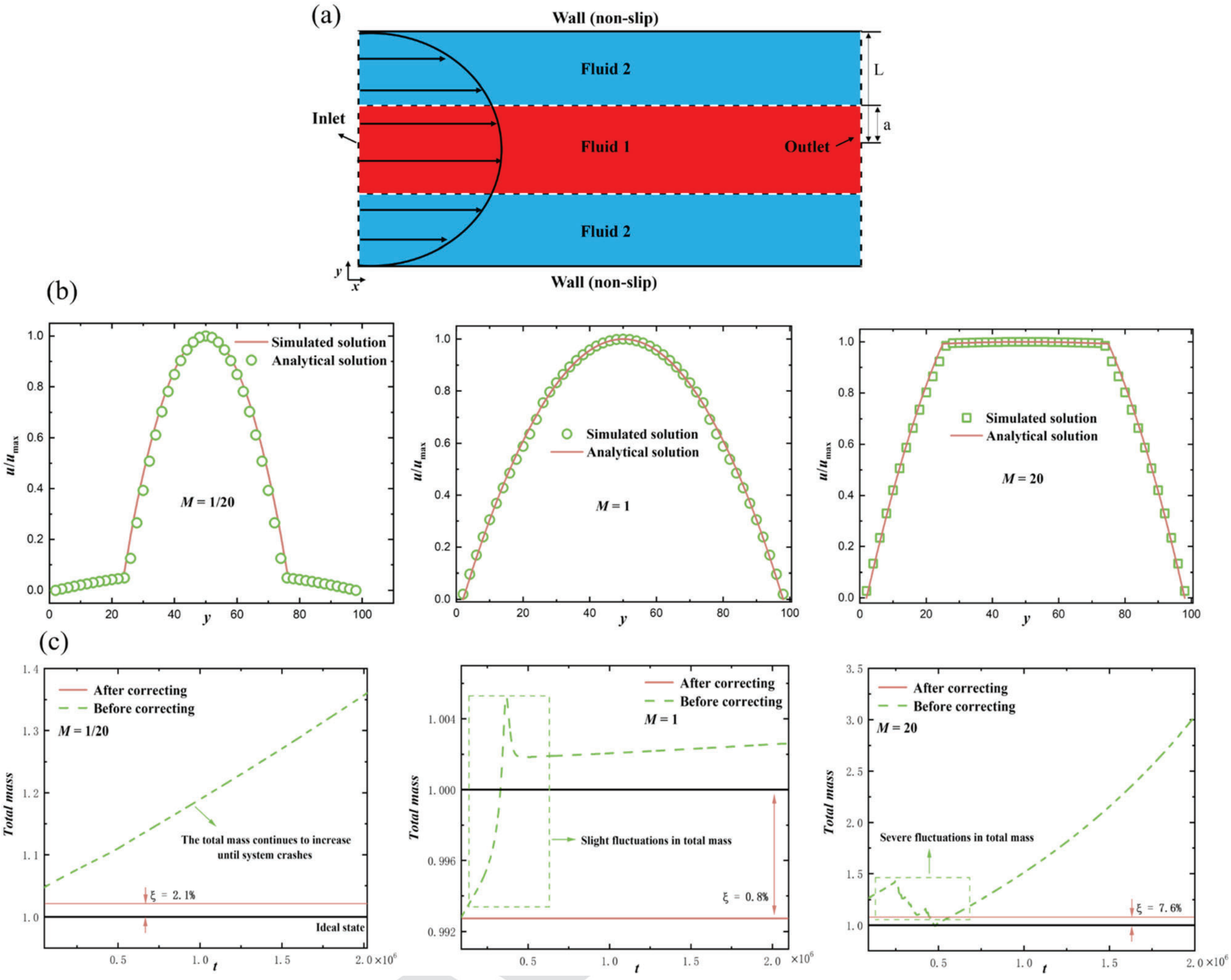

**FIG. 5.** Model and validation of Poiseuille flow at different viscosity ratios, as well as mass conservation performance. (a) Schematic diagrams. (b) Velocity profiles compared with analytical solutions. (c) Time-dependent variation of the total mass in the computational domain.

For the other scene, the computational domain is a $640 \times 320$ grid and the droplet's radius remains unchanged, but its position changes to (70, 160), as shown in Fig. 7(a). Additionally, considering that when the viscosity of the continuous phase fluid is relatively high, the shear effect exerted on the droplet becomes stronger, which can generate greater tangential velocity. Therefore, the viscosity of fluid 2 was adjusted to 1.67. As shown in Fig. 7(b), it is observed that the predictions given by both the corrected scheme and original scheme are nearly identical, with non-visible differences observed, as it was similarly observed in the previous case study. When the droplet approaches the obstacle, it will be deformed due to the increased pressure, and the repulsive force pushes the droplet to move downwards and eventually flow out of the boundary. The above-mentioned two verification cases fully demonstrate that the corrected scheme retains the advantages of the original scheme while ensuring mass conservation.

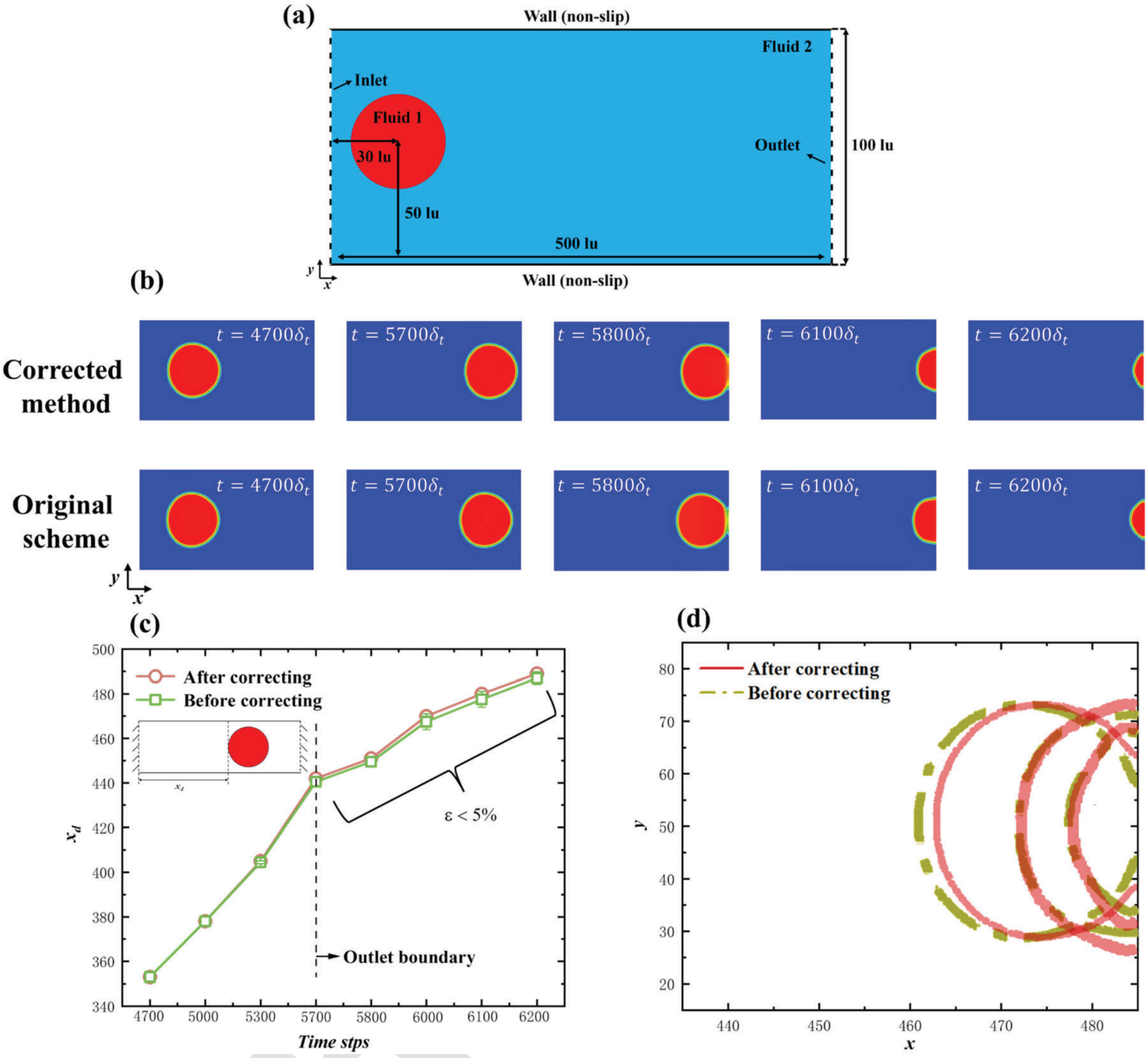

FIG. 6. Model and droplet migration process. (a) Schematic diagrams. (b) Transient motion of the droplet. (c) The quantitative instantaneous position of droplets at different times. (d) The degree of phase interface deformation at the outlet.

In order to further verify the applicability of the corrected method in 3D situations, this paper extends to 3D with a computational domain of $500 \times 100 \times 100$ and compares the situation of mass conservation of the fluid system before and after correction, as shown in Fig. 8. Before flowing through the outlet boundary, the motion posture of the droplet is almost identical under the action of two methods. However, when the droplet passing through the outlet boundary, there is a difference in the results of the two methods, and the source of this difference can be seen from Fig. 8(b). When the droplet passing through the outlet boundary, the original outflow scheme induces significant fluctuations in the global mass variation of the fluid system. The corrected method can correct the velocity of the outlet boundary to suppress the mass deviation of the fluid system, resulting in a change in the droplet shape compared to the original scheme. Overall, the above-mentioned validation further reflects the shortcomings of the original outflow scheme in 3D applications and the necessity of the proposed corrected method in this paper.

## D. Droplet generation in T-shaped and co-flow devices

The fundamental validation cases presented in Secs. III A–III C confirm the accuracy and robustness of the corrected methods. Building upon these results, the present section extends the application of the corrected methods to droplet generation systems—specifically, T-shaped and co-flow configurations—in order to evaluate its performance under more practical and complex operating conditions. Figures 9(a) and 10(a) demonstrate a schematic diagram of a T-shaped and co-flow microchannel, respectively. A T-shaped channel is considered in microscale, having two inlets and one outlet. The size parameters of this structure are consistent with those of Shi *et al.*[43] The fluid 2, as the continuous phase, is injected through the main channel. While the fluid 1, as the dispersed phase, is injected through the lateral channel. In the simulations, the inlet and outlet boundaries are the corrected NEQ and outflow scheme, respectively. At the position where the fluids meet the solid walls, the halfway bounce-back boundary condition scheme is employed to achieve the non-slip boundary condition at the solid walls. For co-flow structures, this paper refers to the structural design of Chen *et al.*,[44] fluid 1 is injected as a dispersed phase from the middle channel, while fluid 2 is injected as a continuous phase from both sides of the channel simultaneously, and the implementation of other boundary schemes is consistent with that of the T-shaped microchannel.

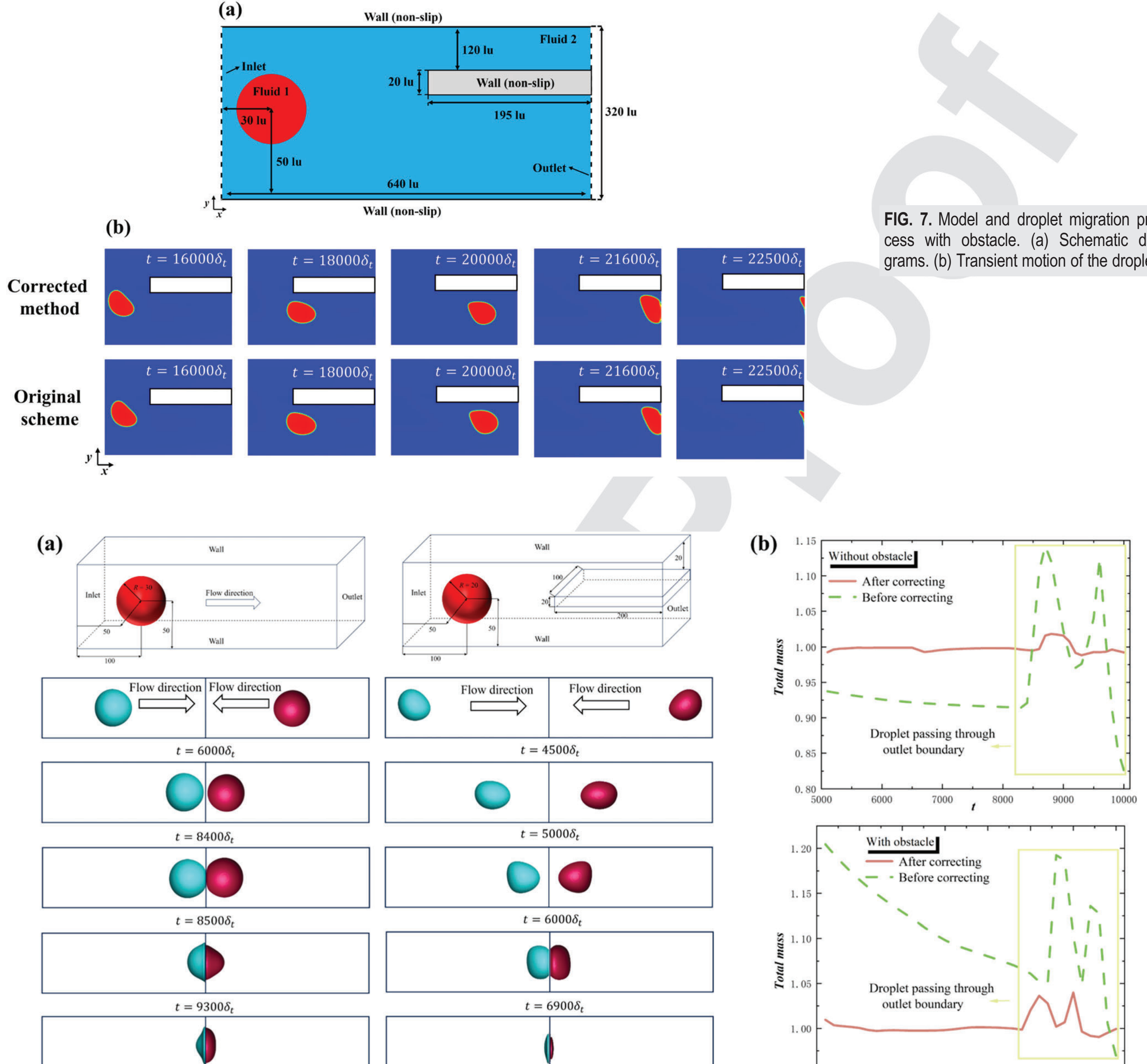

**FIG. 7.** Model and droplet migration process with obstacle. (a) Schematic diagrams. (b) Transient motion of the droplet.

**FIG. 8.** Model and droplet migration process. (a) Schematic diagrams and transient motion of the droplet. (b) Global mass variation of fluid systems.

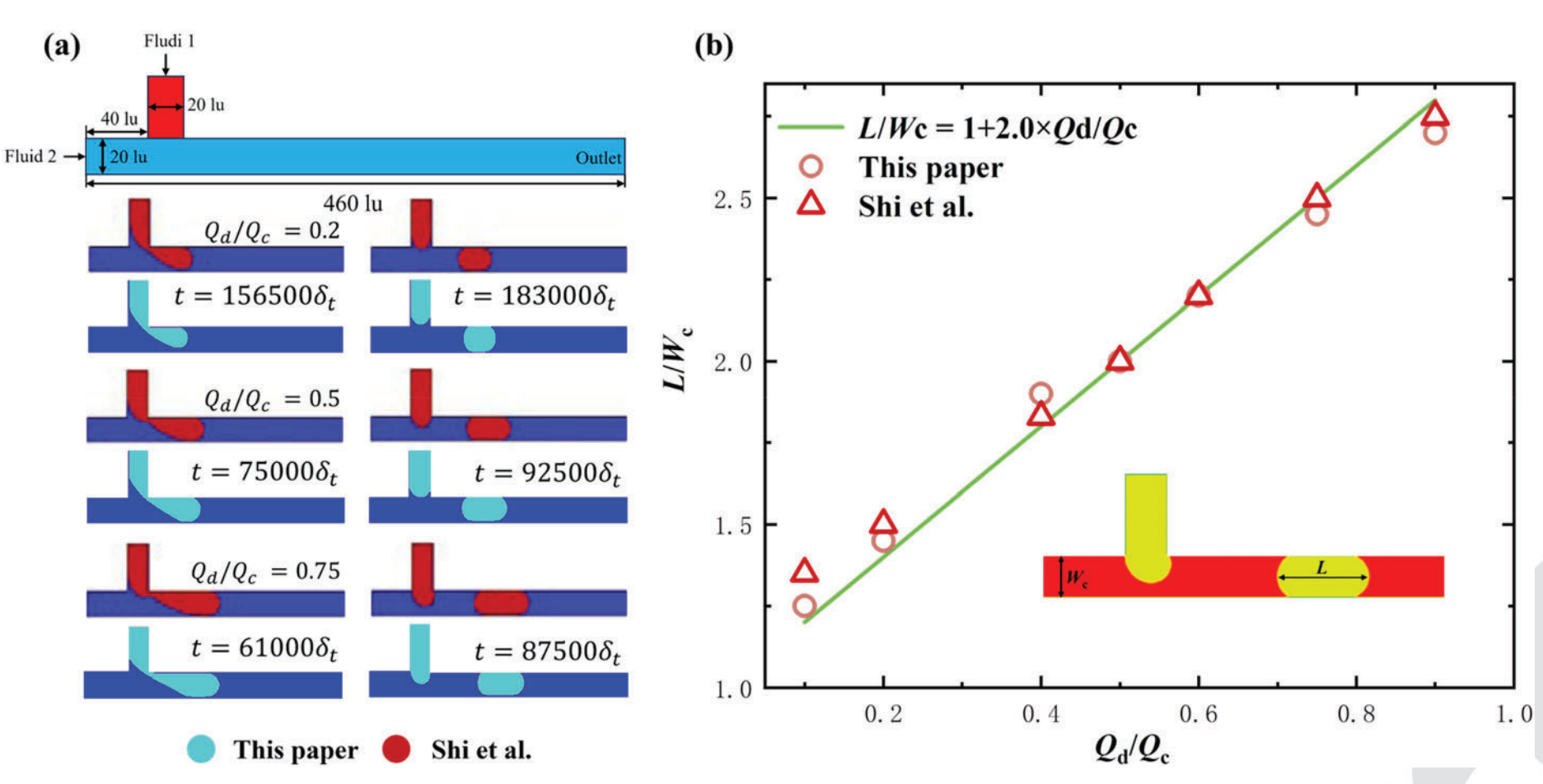

FIG. 9. Model and its validation. (a) Schematic diagrams and droplet formation process. (b) plug length plotted as a function of flow rate ratio $Q_d/Q_c$ compared with Shi *et al.*[43]

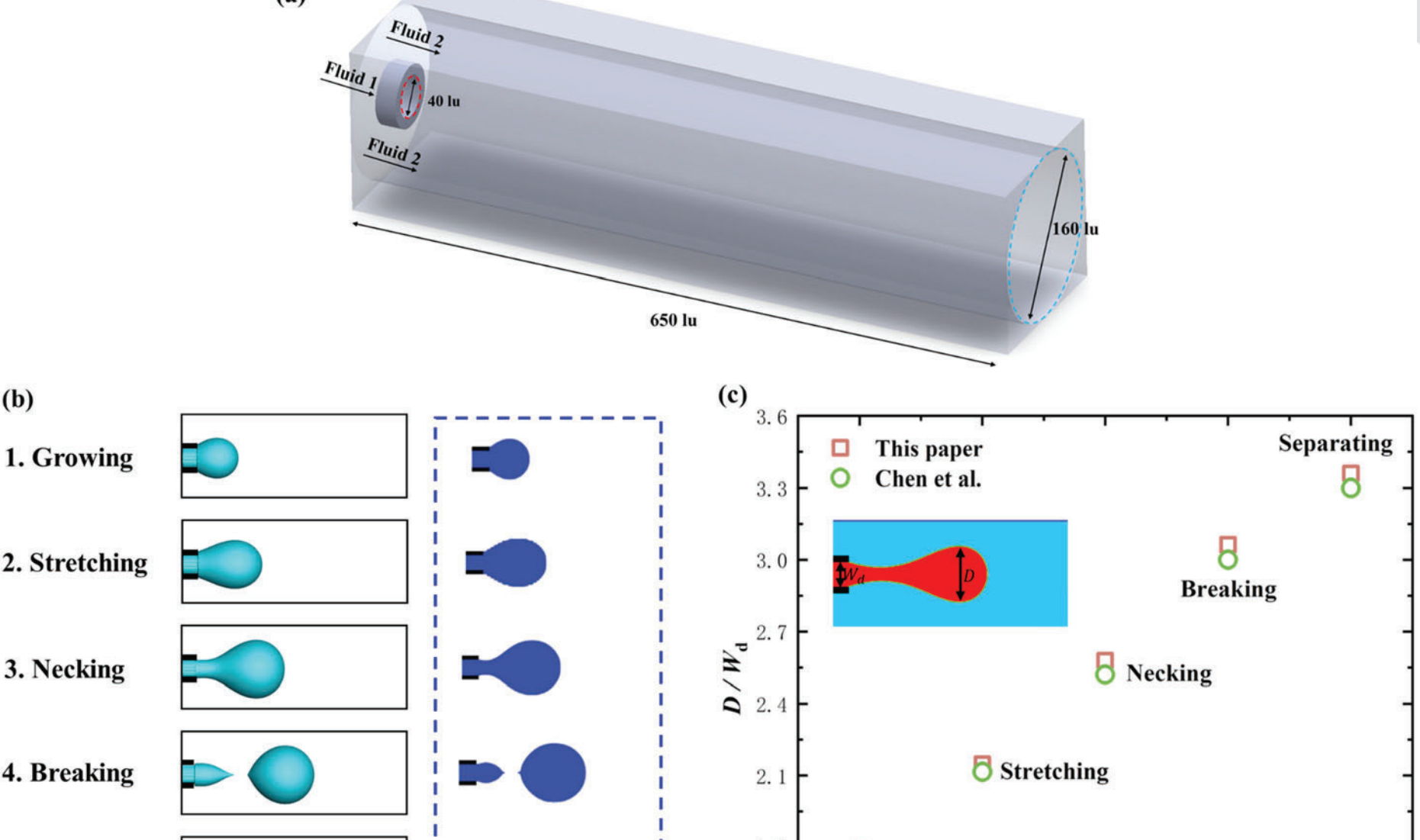

FIG. 10. Model and its validation. (a) Schematic diagrams. (b) Droplet formation process and (c) dimensionless size compared with Chen *et al.*[44]

First, this paper compared the results of the plug flow simulated by Shi *et al.*[43] and the physical properties and wall wettability of the fluid are also consistent with it, as shown in Fig. 9(a). As the droplet forms, it approaches complete occlusion of the main channel's cross-sectional area. Under the shear exerted by the continuous phase, a liquid neck is formed at the T-junction and subsequently undergoes pinch-off, leading to the generation of a discrete droplet. Moreover, for the droplets generated in the plug flow mode, their lengths exhibit a linear relationship to the flow rate ratio ($L/W_c = 1 + \alpha \cdot Q_d/Q_c$). Therefore, this paper observed the dimensionless length changes of droplets at different flow ratios, as shown in Fig. 9(b). The results indicate that the length of separated droplets increases with the increase in dispersed phase flow rate, and the deviation between the simulation results in this paper and the results of Shi *et al.*[43] is below 5%. This further proves the accuracy of the droplet generation model based on T-shaped microchannel constructed in this paper. Subsequently, this paper also validated the droplet generation system based on co-flow structure and compared it with the results of Chen *et al.*,[44] as shown in Fig. 10(b). In the initial growth stage, the droplet volume increases due to the continuous addition of the internal fluid, while the interfacial morphology remains spherical under the action of surface tension. As the droplet size further expands, the enhanced shear force induces stretching of the droplet along the flow direction, leading to a gradual transition from a spherical to a pear-shaped profile. Simultaneously, a necking region begins to form between the droplet and the tip. Subsequently, the liquid bridge progressively contracts, fractures, and eventually detaches from the tip. Additionally, this paper further compares the dimensionless size of droplets in the five generation stages, as shown in Fig. 10(c). The dimensionless diameter of the droplets is also highly consistent with the results of Chen *et al.*'s results. The successful validation of the proposed modified method across these complex flow scenarios further corroborates its accuracy and robustness. It should be noted that, in the creation of 3D coaxial flow structures, this paper is based on the inherent mathematical equations of geometric structures and implemented through programing. For example, for internal and external pipelines, the mathematical equation description is the cylinder equation, as shown in Eq. (28). Subsequently, in programing, the interior of the cylinder can be set as a fluid domain, while the exterior is a solid domain that does not participate in the calculation. In addition, due to the enormous computational resources required for 3D computing, if a single core serial approach is still used, the time cost will be enormous. Therefore, this paper adopts the Open Multi-Processing (OpenMP) parallel algorithm based on multi-threading to accelerate the computing process, and at the end of this paper, a specific algorithm step is also shown as an example to illustrate the implementation details, as shown in Appendix D

$$\begin{cases} y^2 + z^2 \leq R^2, \\ 0 \leq x \leq H. \end{cases} \tag{28}$$

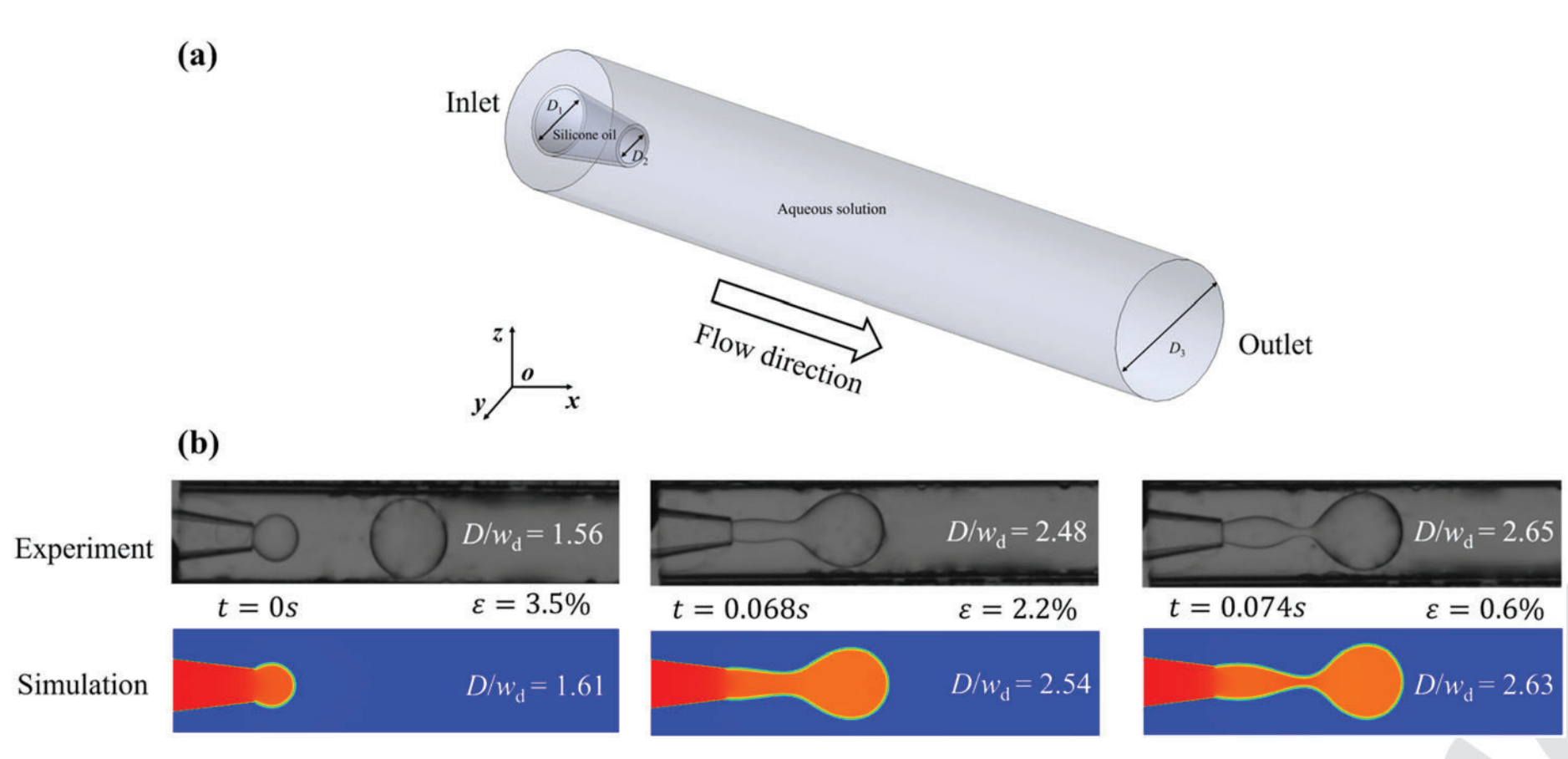

**FIG. 11.** Comparison of simulation and experimental results on focused structure. (a) Schematic diagrams. (b) Droplet diameter-to-tube's width ($D/w_d$) at different time.

$x$, $y$, and $z$ are, respectively, the horizontal, vertical, and depth axis in the Cartesian coordinate system. $R$ and $H$ are cylinder radius and height, respectively.

In the method of generating droplets using coaxial flow structures, except the non-focused structure mentioned above, there is also a focused structure, as shown in Fig. 11(a). Compared to non-focused structures, focused structures can generate a larger range of droplet sizes and improve monodispersity, making it widely used in fields such as polymerase chain reaction and synthesis of high-performance microsphere materials.[45] However, in numerical simulations, due to the fact that the focused structure can cause a stronger pressure drop and higher flow velocity, the numerical stability at the inlet is affected to some extent. Therefore, to further test the applicability of the corrected scheme proposed in this paper, a focused coaxial flow structure was used to generate droplets with a jetting pattern ($Ca_d = 0.38$ and $Ca_c = 0.003$), and the simulation results were compared with the experiment results, as shown in Fig. 11(b). The instruments and fluid properties used in the experiment are all visible in Appendix E. Compared with the experimental results, the simulation reproduced generation process at different times with an average absolute deviation small than 5%, which further verifies the accuracy of the proposed modified boundary method in this paper.

## IV. CONCLUSION

The pseudopotential method is widely used in the field of droplet dynamics due to its clear concept and simple implementation. However, there is little research on the modeling of droplet generation using this method, mainly due to the influence of spurious currents on interface formation and breakup, as well as the impact of inlet/outlet boundary configuration on mass conservation. Therefore, the main contribution of this paper lies in three aspects of enhancement for the open boundary treatment of immiscible pseudopotential models:

(1) Precise boundary implementation: By introducing distribution function correction coefficients to correct the distribution function of the inlet boundary, the error impact caused by interpolation in the original non-equilibrium extrapolation scheme is effectively reduced.
(2) Ensured mass conservation: Based on real-time import and export mass flow rates, a velocity correction coefficient is introduced to correct the velocity at the outlet boundary, achieving a net mass flow rate of zero within the calculation domain and ensuring global mass conservation.
(3) Suppressed spurious currents: According to the kinematic viscosity of the two-phase fluid, the relaxation parameter A that

affects the stability of the model is adjusted to weaken the false velocity at the phase interface.

Validation through four benchmark cases—Laplace tests and Taylor deformation, two-phase Poiseuille flow, droplet migration in microchannels, and droplet generation in T-shaped and co-flow devices—demonstrates the versatility and accuracy of our extended methods. Quantitative comparisons with analytical solutions and prior studies reveal relative errors below 5%, affirming its reliability for diverse multiphase flow scenarios. This work establishes a foundation for high-fidelity simulations of droplet generation phenomena in pseudopotential LBM frameworks, with potential applications in energy systems, biomedical engineering, and environmental fluid dynamics.

## ACKNOWLEDGMENTS

The authors acknowledge the financial support of the National Natural Science Foundation of China (No. 52576065), the Natural Science Foundation of Guangdong Province (No. 2024A1515012393), and the Guangzhou Science and Technology Program—Science Elite Pioneering Project (No. 2025A04J7055).

## AUTHOR DECLARATIONS

### Conflict of Interest

The authors have no conflicts to disclose.

### Author Contributions

**Yizhong Chen:** Methodology (equal); Software (equal); Validation (equal); Writing – original draft (equal); Writing – review & editing (equal). **Huajie Pan:** Methodology (equal); Software (equal). **Zhibin Wang:** Conceptualization (equal); Data curation (equal); Formal analysis (equal); Funding acquisition (equal); Writing – review & editing (equal). **Dongliang Li:** Writing – review & editing (equal). **Ying Chen:** Funding acquisition (equal); Project administration (equal); Writing – review & editing (equal).

## DATA AVAILABILITY

The data that support the findings of this study are available within the article.

## APPENDIX A: ALGORITHM

**ALGORITHM:** Corrected open boundary framework.

// Inlet boundary
// Calculate the corrected distribution function at the inlet boundary using the corrected coefficient
1: for each node $(x_1, y)$ do
2: if (the node located at $(x_1, y)$ is a fluid node) then
3: calculating $f_i^{k(eq)}$ $(x_1, y)$ and $f_i^{k(eq)}$ $(x_2, y)$ using Eq. (6)
4: Calculate corrected coefficient $\alpha$ based on the given velocity at the inlet boundary using Eq. (19)
5: calculating $f_i^{k}$ $(x_1, y)$ using Eqs. (13) and (15)
6: end if
7: end for
// Outlet boundary
// Calculate the corrected velocity at the outlet boundary according to the conservation of mass
8: for each node $(x_{N-1}, y)$ do
9: if (the node located at $(x_{N-1}, y)$ is a fluid node) then
10: calculating $\chi$ using Eq. (23)
11: end if
12: end for
13: for each node $(x_{N-1}, y)$ do
14: if (the node located at $(x_{N-1}, y)$ is a fluid node) then
15: calculating corrected velocity using Eq. (22)
16: end if
17: end for

Source code link: shixuemozun/A-Corrected-Open-Boundary-Framework-for-Lattice-Boltzmann-Immiscible-Pseudopotential-Models

## APPENDIX B: DETAILED DESCRIPTION OF BOUNDARY SCHEME

For the treatment of non-slip conditions, this paper adopts half-bounce back scheme, represented by Eq. (B1), which can effectively ensure the conservation of mass in the simulation system; for cyclic boundaries, as shown in Eq. (B2), this scheme is mainly used

**TABLE I.** Dimensionless conversion (example for the case in Sec. III C).

| Physical property | SI units | Conversion factor | Lattice units |
|---|---|---|---|
| Length | $5 \times 10^{-4}$ m | $C_l = \frac{5 \times 10^{-4}}{500} = 10^{-6}$ | 500 lu |
| Density | $10^3$ kg/m$^3$ | $C_m = \frac{10^3}{1.0} \cdot C_l^3 = 10^{-15}$ | $\rho_1 = \rho_2 = 1.0$ mu/lu$^3$ |
| Time step | $1.7 \times 10^{-7}$ s | $C_t = \frac{C_l^2}{C_\upsilon} = 1.7 \times 10^{-7}$ | 1 ts |
| Kinematic viscosity | $10^{-6}$ m$^2$/s | $C_\upsilon = \frac{10^{-6}}{0.17} = 5.88 \times 10^{-6}$ | $\upsilon_1 = \upsilon_2 = 0.17$ lu$^2$/ts |
| Surface tension | $13.5 \times 10^{-3}$ kg/s$^2$ | $C_\gamma = \frac{C_m}{C_t^2} = 0.035$ | $\gamma = 0.385$ mu/ts$^2$ |
| Velocity | $2.94 \times 10^{-1}$ m/s | $C_v = \frac{C_l}{C_t} = 5.88$ | $5 \times 10^{-2}$ lu/ts |

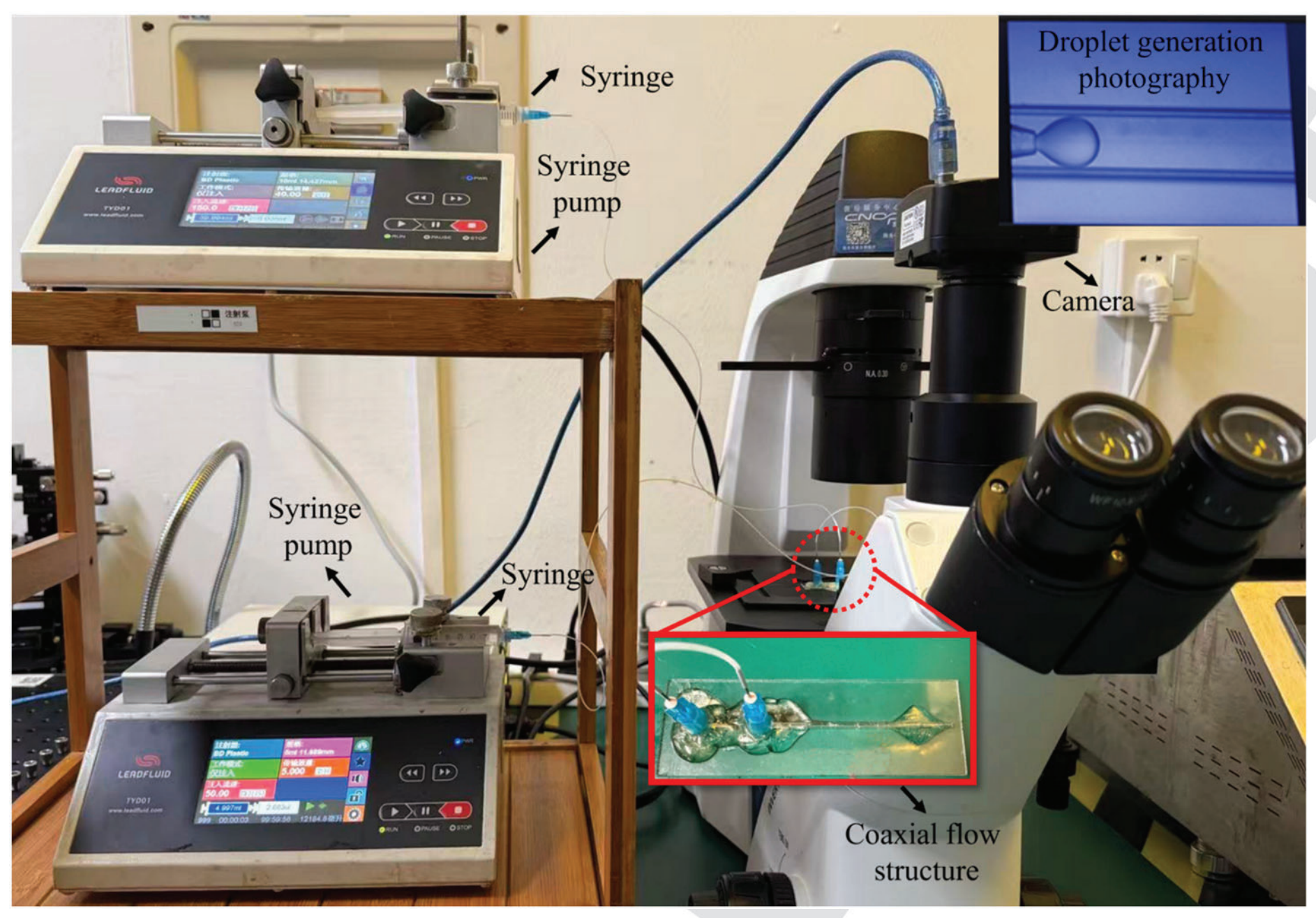

**FIG. 12.** Experiment on droplet generation based on coaxial flow structure.

to handle infinitely long channels and can also ensure strict conservation of mass

$$f_{i_}^{k}(\boldsymbol{r}_b, t+\Delta t) = f_i^{*,k}(\boldsymbol{r}_b, t). \tag{B1}$$

where $\boldsymbol{r}_b$ is the fluid boundary node, $f_i^{*,k}$ is distribution function after collision, $i_$ and $i$ represent opposite discrete velocity directions, respectively

$$f_{1,5,8}^{*,k}(r_{b_r}, t) = f_{1,5,8}^{k}(r_{b_l}, t+\Delta t), \tag{B2a}$$

$$f_{3,6,7}^{*,k}(r_{b_l}, t) = f_{3,6,7}^{k}(r_{b_r}, t+\Delta t), \tag{B2b}$$

$$f_{2,5,6}^{*,k}(r_{b_u}, t) = f_{2,5,6}^{k}(r_{b_d}, t+\Delta t), \tag{B2c}$$

$$f_{4,7,8}^{*,k}(r_{b_d}, t) = f_{4,7,8}^{k}(r_{b_u}, t+\Delta t). \tag{B2d}$$

$\boldsymbol{r}_{b_u}$, $\boldsymbol{r}_{b_d}$, $\boldsymbol{r}_{b_l}$, and $\boldsymbol{r}_{b_r}$, respectively, represent the upper, down, left, and right boundaries.

## APPENDIX C: DIMENSIONAL CONVERSION

Dimensionless conversion (example for the case in Sec. III C)

## APPENDIX D: IMPLEMENTATION OF OPENMP

Parallel computing OpenMP.

```
// Calculate macroscopic quantities for 3D
// Parallel statements which is set before for-loop. num_omp is the number of parallel threads, for_num is the number of merged loops, and the traversal indices i, j and z of the array are set as private variables to avoid concurrent access. In this example, num_omp = 64, for_num = 3
#pragma omp parallel for num_threads(num_omp) collapse(for_num) private(i, j, z)
for (i = 3; i < NX - 2; i++)
{
for (j = 2; j < NY - 1; j++)
{
for (z = 2; z < NZ - 1; z++)
{
Calculate density and velocity using Eqs. (7) and (8)
}
}
}
```

## APPENDIX E: EXPERIMENTAL DETAILS

In the experiment of droplet generation, the instruments used include an syringe pump, syringe, microscope, and camera, as shown in Fig. 12. Inject the dispersed phase and continuous phase into the coaxial flow structure separately through an injection pump to complete the droplet generation step. Subsequently, the camera will capture different moments during the droplet generation process and display them on the computer screen. The detail information of experimental instrument, geometric dimensions, and fluid properties parameters used in this experiment are listed in Table II.

TABLE II. Detailed information about equipment and fluid properties.

| Category | Name | | Model no./ Specifications | |
|---|---|---|---|---|
| Instrument | Syringe pump | | TYDO1-01 | |
| | Syringe | | 60 ml | |
| | Microscope | | BDS400 | |
| | Camera | | CCD TP510 | |
| Consumables | Glass capillary | | D1: 0.2 mm<br>D2: 0.17 mm<br>D3: 0.6 mm | |
| Fluids | | Density (kg/m3) | Viscosity (Pa s) | Surface tension (mN/m) |
| | 50 cSt Dimethyl Silicone Oil | 960 | 0.0434 | 13.5 |
| | 5% (w/w) Tween-20 aqueous solution | 996 | $9.8 \times 10^{-4}$ | |

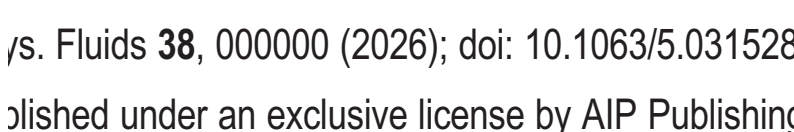